\documentclass[conference,a4paper]{IEEEtran}
\IEEEoverridecommandlockouts 
\usepackage{cite}
\usepackage{dsfont}
\ifCLASSINFOpdf
   \usepackage[pdftex]{graphicx}
   \graphicspath{{./Images/}}
   \DeclareGraphicsExtensions{.pdf,.jpeg,.png}
\else
\fi

\usepackage[cmex10]{amsmath}

\usepackage[hidelinks]{hyperref}
\usepackage{multirow}
\usepackage{booktabs,colortbl}

\usepackage{cite}
\usepackage{psfrag}
\usepackage[utf8]{inputenc}
\usepackage{amsmath,amsfonts,amsbsy,amssymb}
\usepackage{mathabx}
\usepackage{mathrsfs}
\usepackage[nolist]{acronym}
\usepackage{tabularx}
\usepackage{amssymb}
\usepackage{amsmath}
\usepackage{multirow}
\usepackage{wasysym}
\usepackage{multirow}
\usepackage{float}
\usepackage{color}
\usepackage{pifont}
\begin{document}
\title{Decision Error Probability in a Two-stage\\ Communication Network for Smart Grids\\ with Imperfect Data Links}

\author{%

\IEEEauthorblockN{Iran Ramezanipour, Mauricio C. Tomé, Pedro H. J. Nardelli, Hirley Alves}
\IEEEauthorblockA{Centre for Wireless Communications (CWC)\\ 
	University of Oulu, Finland\\
Contact: [iramezan,decastrotome,nardelli,halves]@ee.oulu.fi}

%

\thanks{ This work is partly funded by Finnish Academy and CNPq/Brazil (n.490235/2012-3) as part of the joint project SUSTAIN, and by Strategic Research Council/Aka BC-DC project (n.292854).}%
}

\maketitle

\begin{abstract}
This paper analyzes a scenario where the distribution system operator needs to estimate whether the average power demand in a given period is above a predetermined threshold using an $1$-bit memoryless scheme.
Specifically, individual smart-meters periodically monitor the average power demand of their respective households to inform the system operator if it is above a predetermined level using only a $1$-bit signal.
The communication link between the meters and the operator occurs in two hops and is modeled as binary symmetric channels. 
The first hop connects individual smart meters to their corresponding aggregator, while the second connects different aggregators to the system operator.
A decision about the power demand also happens in two stages based on the received information bit.
We consider here three decision rules: AND, OR and MAJORITY.
Our analytical results indicate the circumstances (i.e. how frequent the meters experience the consumption above the defined threshold) and the design setting (i.e. decision rules) that a low error probability can be attained. 
We illustrate our approach with numerical results from actual daily consumption from $12$ households and $3$ aggregators.
\end{abstract}

\vspace{2ex}


\section{Introduction}
The basic structure of the electrical power grids have been the same for the past century \cite{bush2014smart}.
Several problems such as more voltage sags, blackouts, and overloads have been raised from using the traditional power grid systems, especially in the past decade as a result of slow response time of devices over the grid~\cite{gao2012survey}.
In addition to that, with the growing size of the population, the demand for electricity and consumption is also increasing~\cite{gungor2011smart}.
This means more appliances and consumers are joining the current power grids which are not designed for handling these large amounts of users. Also, the current power grids contribute greatly to the carbon emission~\cite{farhangi2010path}.
With both economic and environmental aspects in mind, changing the power grid seems inevitable.

The concept of smart grids has been introduced to characterize the modernization of the traditional electrical power grid, empowered by the advances of information and communication technologies~\cite{nardelli2014models}.
Approaching the modernization of power grids in such a way has the potential to mitigate the current energy crisis~\cite{wang2011survey}.
This change shall have consequences to utilities, regulation entities, service providers, technology suppliers, and electricity consumers too~\cite{zaballos2011heterogeneous}: the smart grid is based on a two-way power delivery and frequent communication between different elements of the network~\cite{fan2013smart}.
Looking at the communication side, simple and efficient schemes are very important to guarantee the operation of the power grid without overwhelming the communication network~\cite{sauter2011end,cisco2009internet,nielsen2015can}.

In this paper, we focus on this aspect by analyzing a non-critical application where the system operator needs to estimate whether the average power demand in a given period (e.g. $15$ minutes) of the distribution grid is above a predetermined threshold.
Our goal is to build an efficient communication system with simple and low cost implementation.
To do so, we follow our previous work \cite{nardelli2015average} to build a WSN in two hops such that individual smart meters send to their respective aggregator an $1$-bit message indicate whether the individual average power demand is above a given threshold.
The aggregators then decide about their state based on the received information and send their decision as an $1$-bit signal to the system operator, which its turn decides if the power demand is above the threshold in the same way.

It is worth saying that the design and implementation of communication networks in smart grids face several challenges.
The main problem is that there are many different ways of building a communication network for smart grids~\cite{gao2012survey,locke2010nist,pullins2009west}; and therefore, the so-called smart grid is built upon several different applications with diverse quality requirements~\cite{kuzlu2014communication}.
For example, several wired and wireless communication technologies such as power-line communications~\cite{galli2011grid}, optical fibers~\cite{ancillotti2013role}, IEEE 802.11 based wireless LAN, IEEE 802.16 based WiMAX,3G/4G cellular and ZigBee based on IEEE 802.15 are currently being used in smart grids communication technologies~\cite{parikh2010opportunities}.

\begin{figure*}[!t]
	\centering
	\includegraphics[width=1.5\columnwidth]{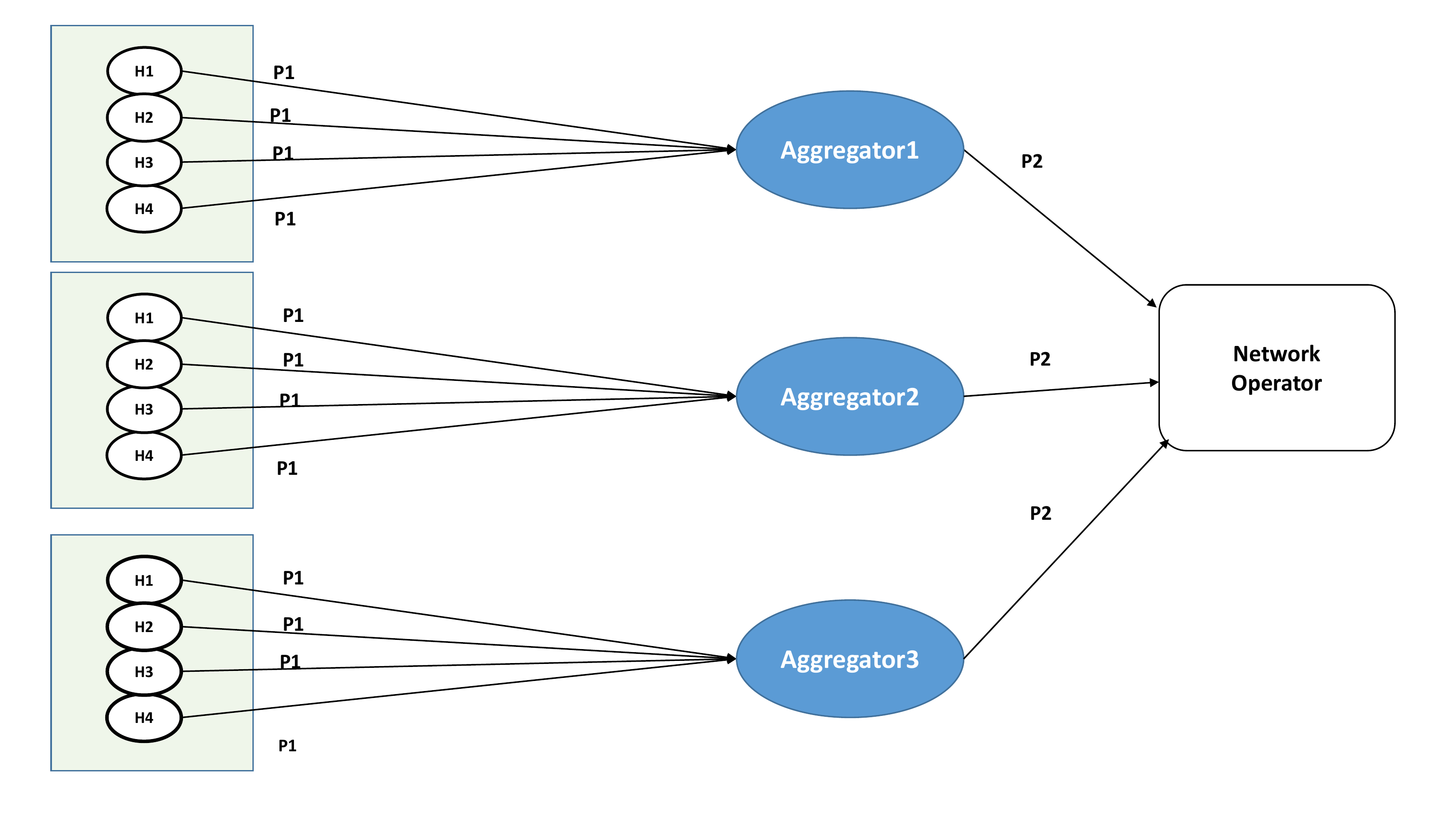}
	\caption{Example of the scenario studied in this in this paper, considering $K=3$ and $N_i = 4$ for all $i$, resulting in $12$ smart meters total.
	Smart meters monitor the average power demand in order to determine its binary state $\theta_{i,j}[n]$ at time $t_n$. The meters are associated with $3$ aggregators that decide their state $\theta_i[n]$ based on the inputs from $4$ household data. The aggregators then send their  state $\theta_i[n]$ to the system operator that will decide about the global state $\theta[n]$. The communication channel is modeled as BSC with probabilities  $p_1$ and $p_2$. }
	\label{fig:schematic}
	\vspace{-2ex}
\end{figure*}

In our case, we choose to build a generic communication model for non-critical applications in the distribution level of the power grid.
In this way, our approach does not focus on high reliability or low latency, but rather on a cheap way to estimate the average power demand without harming the communication network with huge amounts of data (e.g. \cite{nielsen2015can}).
We consider a case where the smart-meters inform the aggregator whether their average power demand in predetermined time periods is above or below a given threshold. 
Aggregators process the received information using AND, OR and MAJORITY (memoryless) logical operations and send the processed information to the 
the system operator.
The system operator then decides based on the same logical operation if the aggregate average power demand is above or below the threshold.
We test our approach using 12 daily demand profiles taken from the database ``The Reference Energy Disaggregation Data Set'' (REDD) \cite{REDD}. Thus, as depicted in Fig.\ref{fig:schematic}, every 4 houses connect to one aggregator, therefore we assume 3 aggregators besides the network operator.

The rest of the paper is divided as follows.
Section \ref{sec:sys} details the system model used here.
In Section \ref{sec:error} we show how we compute the average probability that the system operator decide in success.
Section \ref{sec:result} presents a numerical example of our approach, while Section \ref{sec:final} provides our final discussions.

\section{System model}
\label{sec:sys}
\vspace{-1ex}
Let us assume a network composed by a set $\mathcal{N}_i = \{1,...,N_i\}$ of smart meters of a given group of consumers (\textit{prosumers}) $i$ composed by $N_i$ elements, which are associated with aggregator $i \in \mathcal{N}$ where $\mathcal{N} = \{1,...,K\}$ is the set of aggregators.
Each meter $j \in \mathcal{N}_i$ needs to inform aggregator $i$ in predetermined times $t_n = t + n\tau$ if its individual average power demand, $P_{i,j}(t_n)$, is above or below a given threshold $\gamma$.
Let $\theta_{i,j}[n]$  be the  function that indicates whether $P_{i,j}(t_n) > \gamma$.
%

We assume that smart-meter $j \in \mathcal{N}_i$ sends its state $\theta_{i,j}[n]$ to aggregator $i$ through a binary symmetric channel (BSC) \cite[Ch.7]{proakis2001digital} with error probability $p_1$ (the subscript ``$1$''  indicates the first communication hop). 
Based on such information, aggregator $i$ decides its state $\theta_{i}[n]$ using hard-decision rules AND, OR or MAJORITY from the inputs $\theta_{i,j}[n]$ as follows.

\vspace{1ex}
\begin{itemize}
\itemsep1ex
\item \textbf{AND:} $\theta_i[n]=0$ if at least one $\theta_{i,j}[n]=0$ for $j \in \mathcal{N}_i$. Then, $\theta_i[n]=1$ if all $\theta_{i,j}[n]=1$.
\item \textbf{OR:} $\theta_i[n]=0$ if all $\theta_{i,j}[n]=0$. Then $\theta_i[n]=1$ if at least one $\theta_{i,j}[n]=1$.
\item \textbf{MAJORITY:} As aggregator $i$ has $N_i$ inputs, then $\theta_i[n]=0$ if $\sum\limits_{j \in \mathcal{N}_i} \theta_{i,j}[n]=0 < N_i/2$ and $\theta_i[n]=1$ if $\sum\limits_{j \in \mathcal{N}_i} \theta_{i,j}[n]=0 > N_i/2$. If $\sum\limits_{j \in \mathcal{N}_i} \theta_{i,j}[n]=0 = N_i/2$, then $\theta_i[n]$ is randomly selected with $50\%$ of chance.
\vspace{1ex}
\end{itemize}

Aggregators $i \in \mathcal{N}$ then needs to send its state $\theta_{i}[n]$ to the system operator in a binary symmetric channel with error probability $p_2$.
With the information from all aggregators in hand, the operator similarly proceeds to decide the global state $\theta[n]$ based on AND, OR or MAJORITY logic operations. 
Fig. \ref{fig:schematic} depicts the proposed system model.

\section{Average error probability}
\label{sec:error}
Let $s[n]$ be the binary function denoting whether $\sum\limits_{i \in \mathcal{N}}\sum\limits_{j \in \mathcal{N}_i} \bar{P}_{i,j}(t_n) > P_\textup{th}$.
In this case the value of $s[n]$ indicates the actual state of the network at time $t_n$ and, therefore, this shall be used as the basis of comparison for the communication scheme proposed in Section \ref{sec:sys}.
By doing so, we can define an error event associated with the measurement done at $t_n$ whenever $\theta[n] \neq s[n]$.

As previously discussed, $\theta[n]$ is built to be a simple and cheap estimation of $s[n]$, which error events would still happen even with perfect communication channels.
Including errors in the communication will further increase the uncertainty of the estimation $\theta[n]$.
Herein, we are interested on the average error probability over $n$ such that
\begin{equation}
P_\textup{er} = \dfrac{1}{n_\textup{max}} \sum\limits_{n=1}^{n_\textup{max}} \mathds{1}[\theta[n] \neq s[n]],
\label{eq:av-error}
\end{equation}
where $\mathds{1}[\cdot]$ is the indicator function and $n_\textup{max}$ is the number of measurements considered.

Table \ref{table:1} exemplifies our framework by showing the average power demand of $12$ households considering $6$ measurements\footnote{More details about the data will be provided in the next section.}.
The state $s[n]$ indicates if $\sum_{i,j} \bar{P}_{i,j}(t_n) > P_\textup{th}$, while $\theta[n]$ is the estimation considering the proposed $1$-bit signaling including communication errors.
In this example, $n_\textup{max} = 6$ and $\sum\limits_{n=1}^{n_\textup{max}} \mathds{1}[\theta[n] \neq s[n]] = 2$ (i.e. two error events happened). 
Then, the average error probability is $P_\textup{er} = 2/6 = 33.3\%$.

In the following section, we will provide the numerical results where the average $P_\textup{er}$ is evaluated for different decision rules and channel error probabilities.
  
\begin{table}[t!]
	\centering
		\caption{Example of how to compute the average error probability}
		\label{table:1}
	\begin{tabular}{|c c c c c c|} 
		\hline
		$n$ & $\sum$$\bar{P}_{i,j}(t_n)$ & $P_\textup{th}$ & $s[n]$ & $\theta[n]$ & Error \\ [0.5ex] 
		\hline\hline
		1 & 3956 & 7500 & 0 & 0 & No\\ 
		\hline
		2 & 7843 & 7500 & 1 & 1 & No \\
		\hline
		3 & 11373 & 7500 & 1 & 0 & Yes \\
		\hline
		4 & 7005 & 7500 & 0 & 0 & No\\
		\hline
		5 & 7897 & 7500 & 1 & 1 & No\\
		\hline
		6 & 6353 & 7500& 0 & 1 & Yes\\ 
		\hline
	\end{tabular}

\vspace{-1ex}	
\end{table}

\section{Numerical results}
\label{sec:result}

In this section, we employ the system model and the framework to compute the average error probability presented in the previous sections to assess the performance of the AND, OR and MAJORITY decision rules, as well as the impact of the communication error probability.
To carry out our analysis, we use  ``The Reference Energy Disaggregation Data Set'' (REDD)
database \cite{REDD} to generate a $15$-minute average power demand over a timespan of $24$ hours (one day) for $12$ different daily profiles, yielding $n_\textup{max}=96$.

\begin{figure}[!t]
	\centering
	\includegraphics[width=\columnwidth]{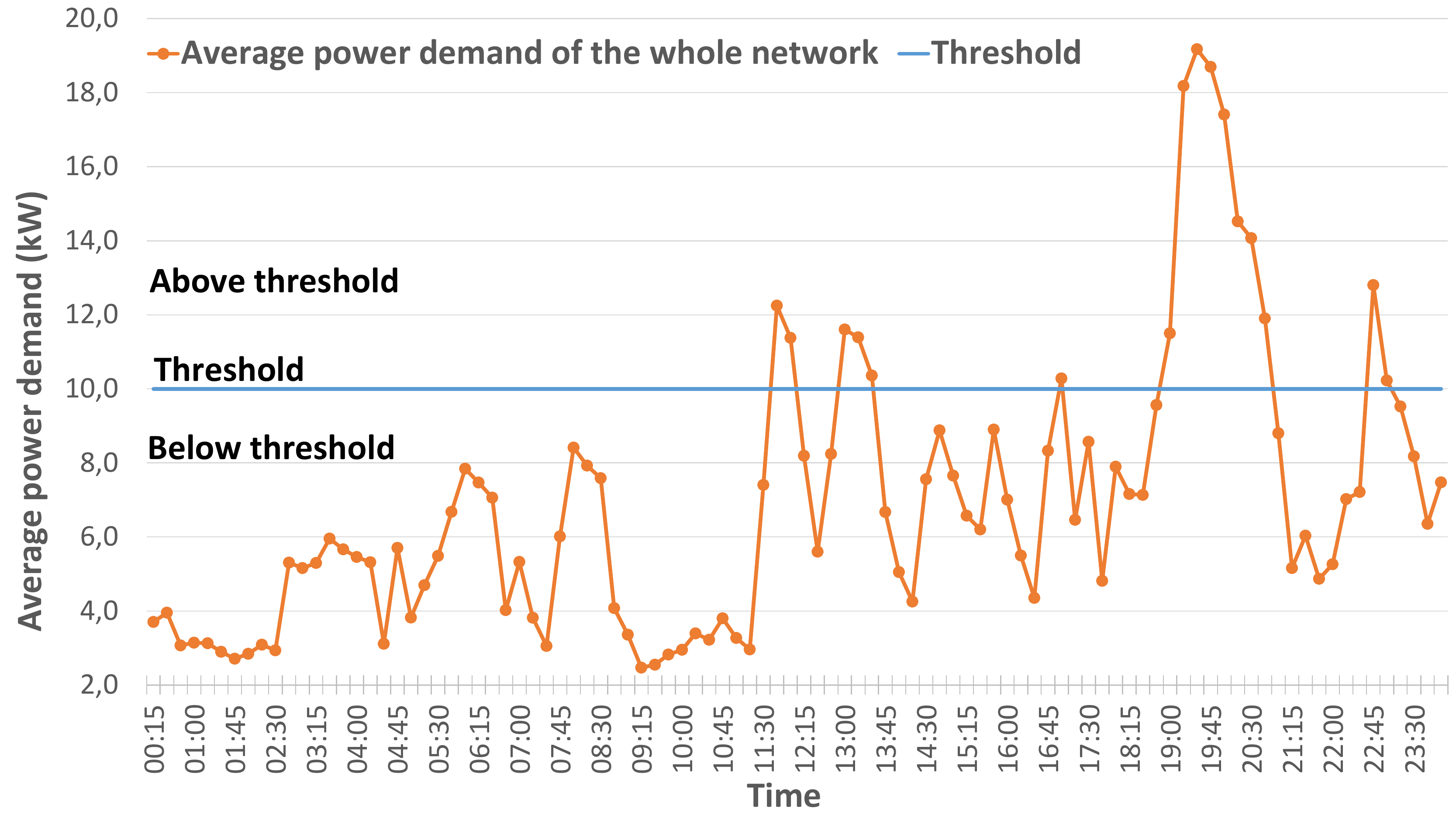}
	
	\caption{An example of the aggregated average power demand $\sum P_{i,j}(t_n)$ and its corresponding threshold $P_\textup{th}$. If $\sum P_{i,j}(t_n) > P_\textup{th}$, then $s[n]=1$; otherwise, $s[n]=0$.}
	\label{fig:total}
\end{figure}

Fig.  \ref{fig:total} exemplifies how the state function $s[n]$ is obtained.
The aggregated average power demand curve is plotted and compared with the system operator threshold $P_\textup{th}$.
If the sample at time $t_n$ is greater than $P_\textup{th}$, then $s[n]=1$; otherwise $s[n]=0$.
As discussed before, $s[n]$ provide the  actual system state that the estimation $\theta[n]$ shall be compared.

\begin{figure}[!t]
	\centering
	\includegraphics[width=\columnwidth]{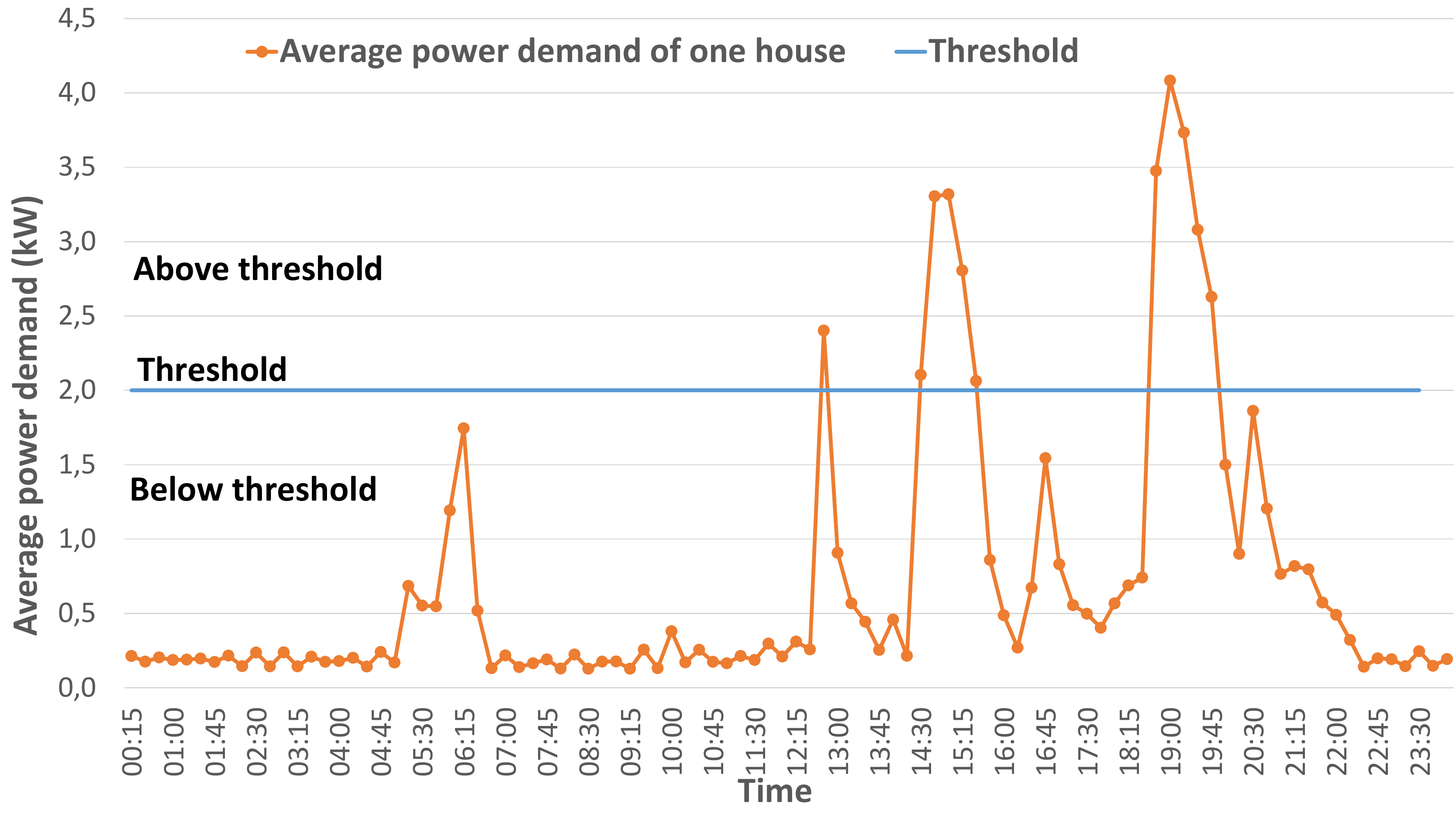}

	\caption{An example of the average power demand for one house $P_{i,j}(t_n)$ in the network and its corresponding threshold $\gamma$. If $P_{i,j}(t_n) > \gamma$, then $\theta_{i,j}[n]=1$; otherwise, $\theta_{i,j}[n]=0$.}
	\label{fig:onehouse}
\end{figure}

\begin{figure}[!t]
	\centering
	\includegraphics[width=\columnwidth]{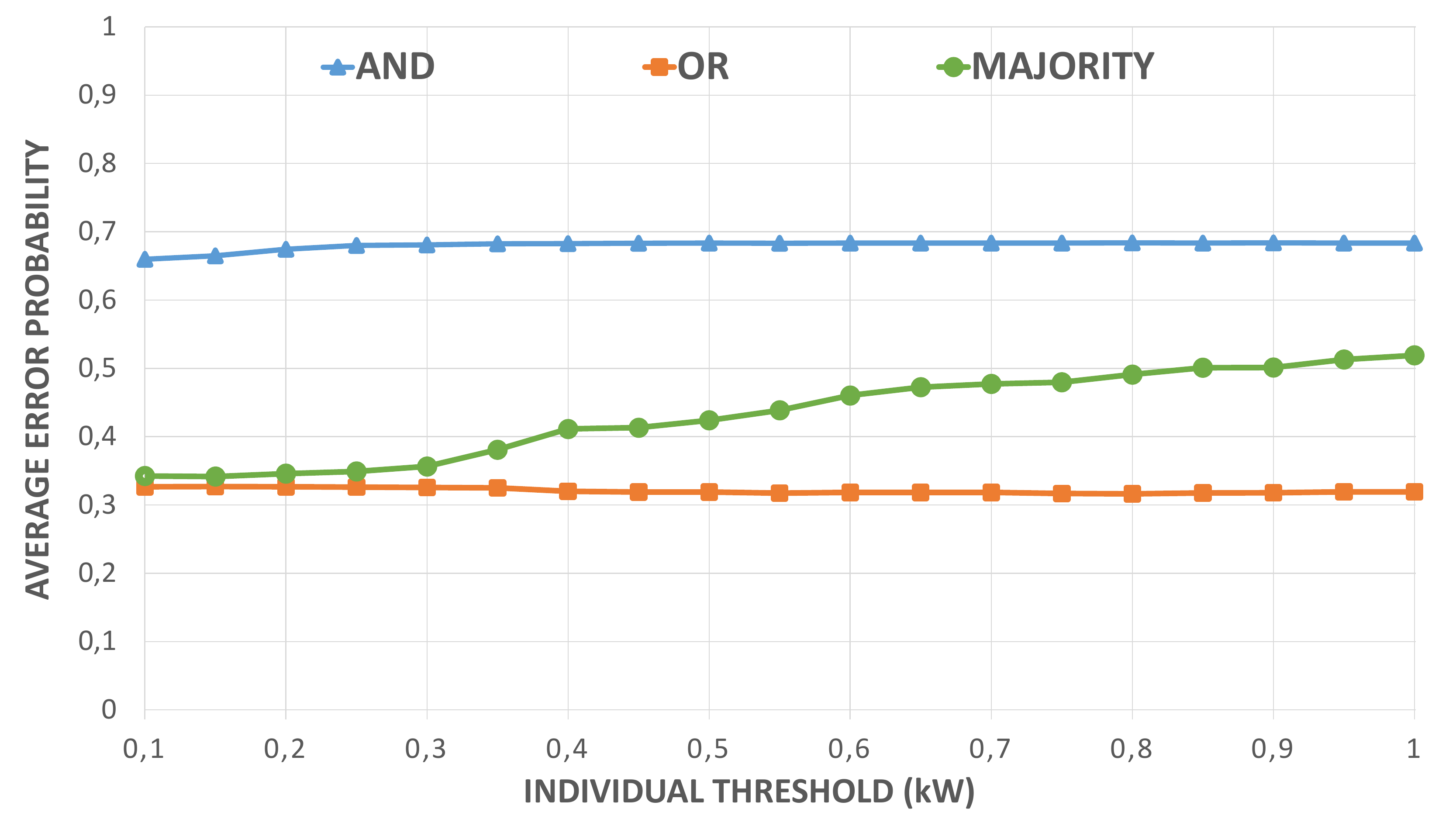}
	\caption{Average Error Probability $P_\textup{er}$ as a function of the individual threshold $\gamma$ assuming AND, OR and MAJORITY decision rules for $p_1=0.2$, $p_2=0.1$ and $P_\textup{th}=5$ kW. Each point is obtained using Monte Carlo simulation ($10^3$  snapshots).}
	\label{fig:5000}
\end{figure}

\begin{figure}[!t]
	\centering
	\includegraphics[width=\columnwidth]{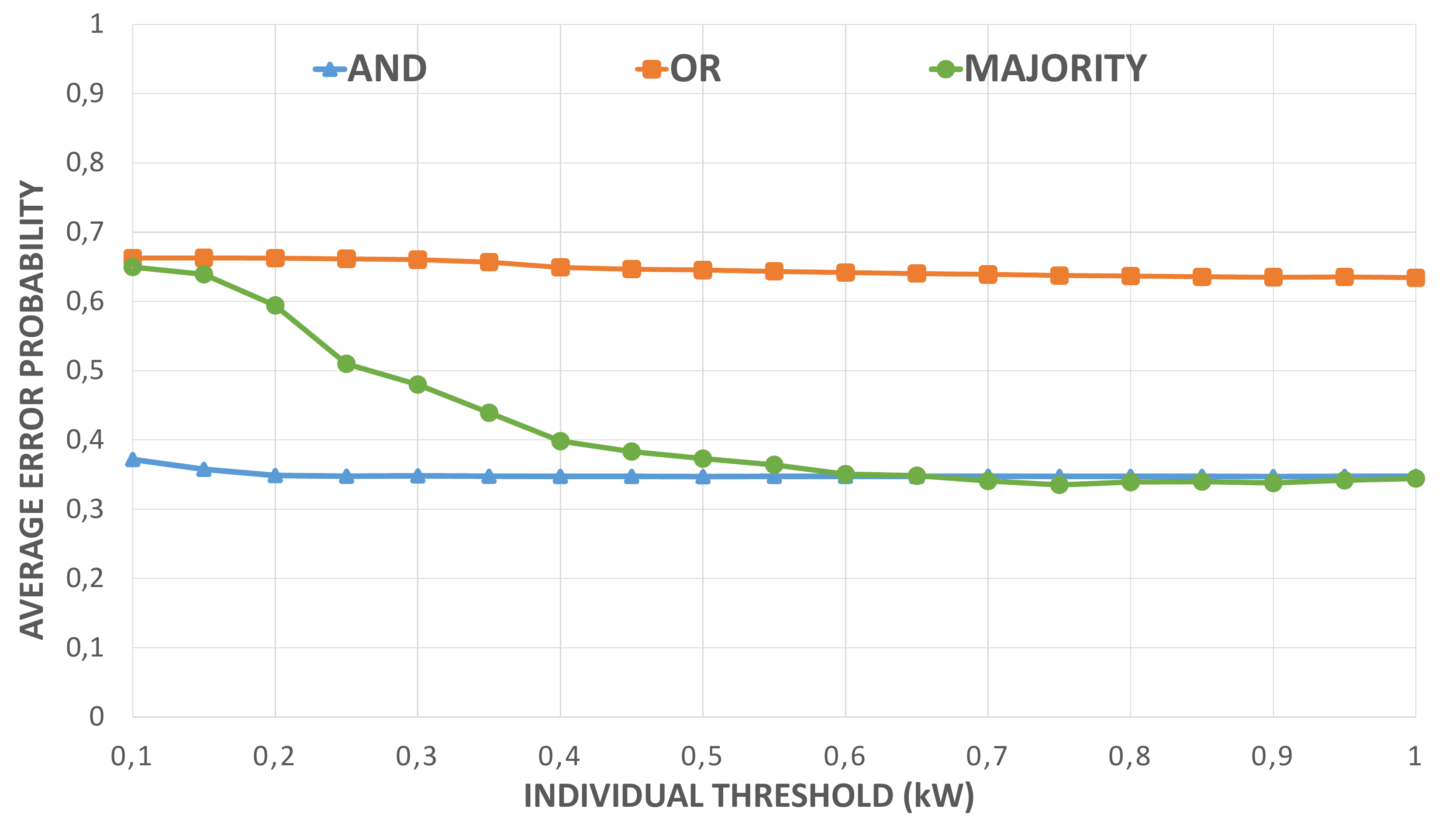}
	\caption{Average Error Probability $P_\textup{er}$ as a function of the individual threshold $\gamma$ assuming AND, OR and MAJORITY decision rules for $p_1=0.2$, $p_2=0.1$ and $P_\textup{th}=7.5$ kW. Each point is obtained using Monte Carlo simulation ($10^3$  snapshots).}
	\label{fig:7500}
	\vspace{1ex}
\end{figure}

\begin{figure}[!t]
	\centering
	\includegraphics[width=\columnwidth]{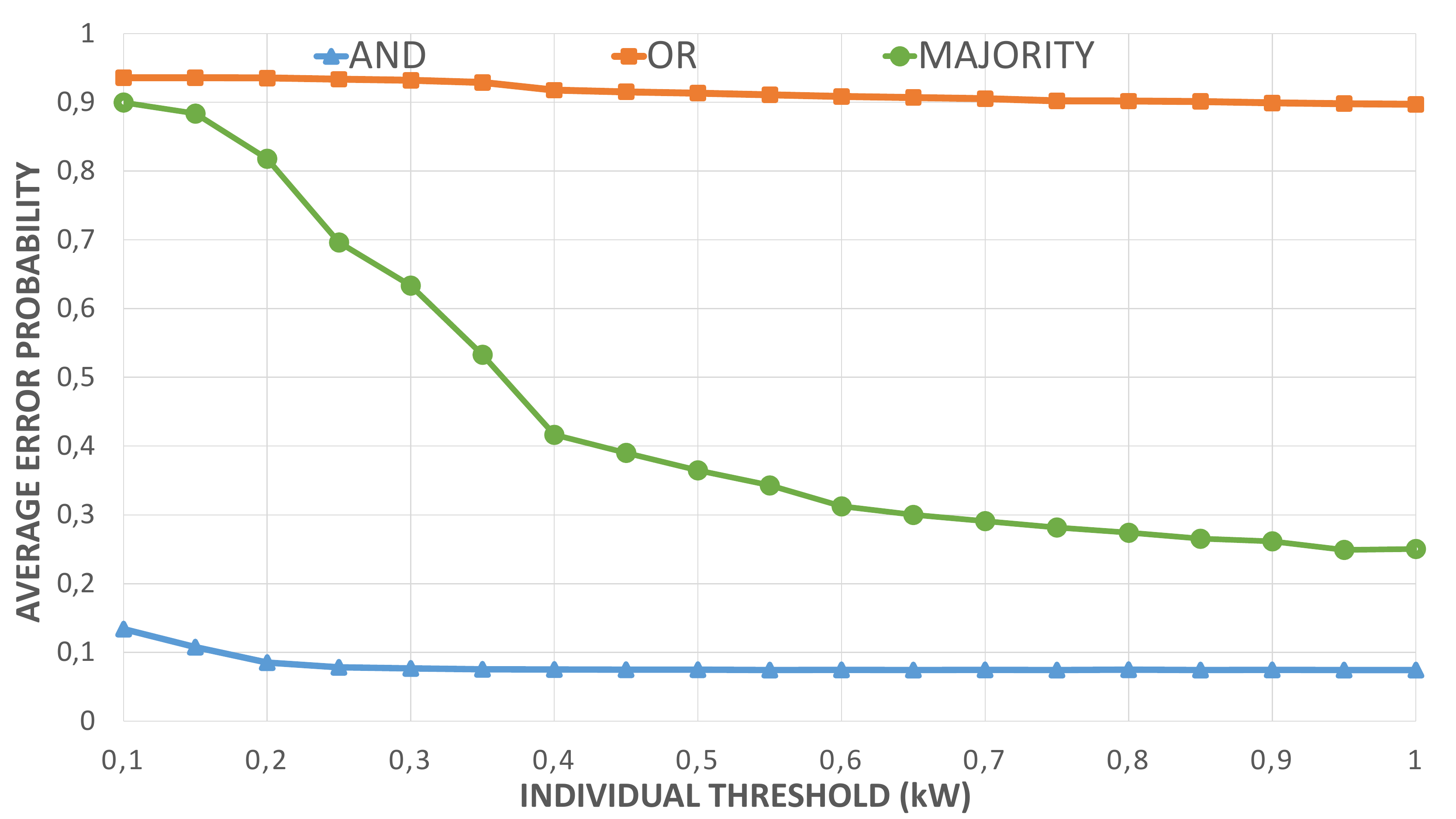}
	\caption{Average Error Probability $P_\textup{er}$ as a function of the individual threshold $\gamma$ assuming AND, OR and MAJORITY decision rules for $p_1=0.2$, $p_2=0.1$ and $P_\textup{th}=12.5$ kW. Each point is obtained using Monte Carlo simulation ($10^3$  snapshots).}
	\label{fig:12500}
\end{figure}

Fig. \ref{fig:onehouse} in its turn show how the individual smart meter code its average power demand $P_{i,j}(t_n)$ into $\theta_{i,j}[n]$.
If the individual demand $P_{i,j}(t_n)$ is above the individual threshold $\gamma$, then $\theta_{i,j}[n]=1$; otherwise, $\theta_{i,j}[n]=0$. 
We proceed similarly with all $12$ households to obtain the states $\theta_{i,j}[n]$ that are the communication system inputs as described in Section \ref{sec:sys}.
Let us focus on the effects of the individual threshold $\gamma$ and the average error probability $P_\textup{er}$, which is given by equation \eqref{eq:av-error}.
Figs. \ref{fig:5000}-\ref{fig:12500} present $P_\textup{er}$ as a function of the individual threshold $\gamma$ for communication error probabilities $p_1=0.2$ and $p_2=0.1$, and the system operator thresholds $P_\textup{th}=5, 7.5$ and $12.5$ kW, respectively.
We consider the threshold $\gamma$ ranges from $0.1$ to $1$ kW.
At the first sight, one can note that both $\gamma$ and $P_\textup{th}$ affects the system reliability.


One can see from Fig. \ref{fig:5000} that when the system threshold $P_\textup{th}$ is low, the scenario with the OR gates works the best.
This is due to the fact that OR gate favors the state ``$1$'' due to its own nature.
Hence, when the threshold is set with a relatively low value, the signal $s[n]=1$ is more frequent.
The individual threshold $\gamma$, however, has little effect on the system performance.
When $P_\textup{th}=5$ kW, the lowest average error probability is about $30\%$.

Increasing the threshold $P_\textup{th}$ modifies this behavior as shown in Figs. \ref{fig:7500} and \ref{fig:12500}: AND gate starts working better with the lower error probability.
Once again, this happens because when the threshold is set higher, the signal $s[n]=1$ becomes more frequent, which favors the performance of AND and results in  error probabilities $P_\textup{er}<10\%$. 
The parameter $\gamma$, once again, has little effect on the error probability for AND and OR.

As for the MAJORITY decision rule, it works most of the times between the other two gates since it does not favor \textit{a priori} any state $s[n]$.
By definition, this rule will choose the state more frequent so the individual threshold $\gamma$ will strongly affect its performance.
In other words, while AND and OR gates respectively induce $\theta[n]=0$ and $\theta[n]=1$, MAJORITY does not induce any state $\theta[n]$.
Therefore, although it can have a worse performance, it can be seen as fairer and better represents the system variations.
This rule is therefore more susceptible to communication errors and variations in the individual thresholds $\gamma$.

\begin{figure}[!t]
	\centering
	\includegraphics[width=\columnwidth]{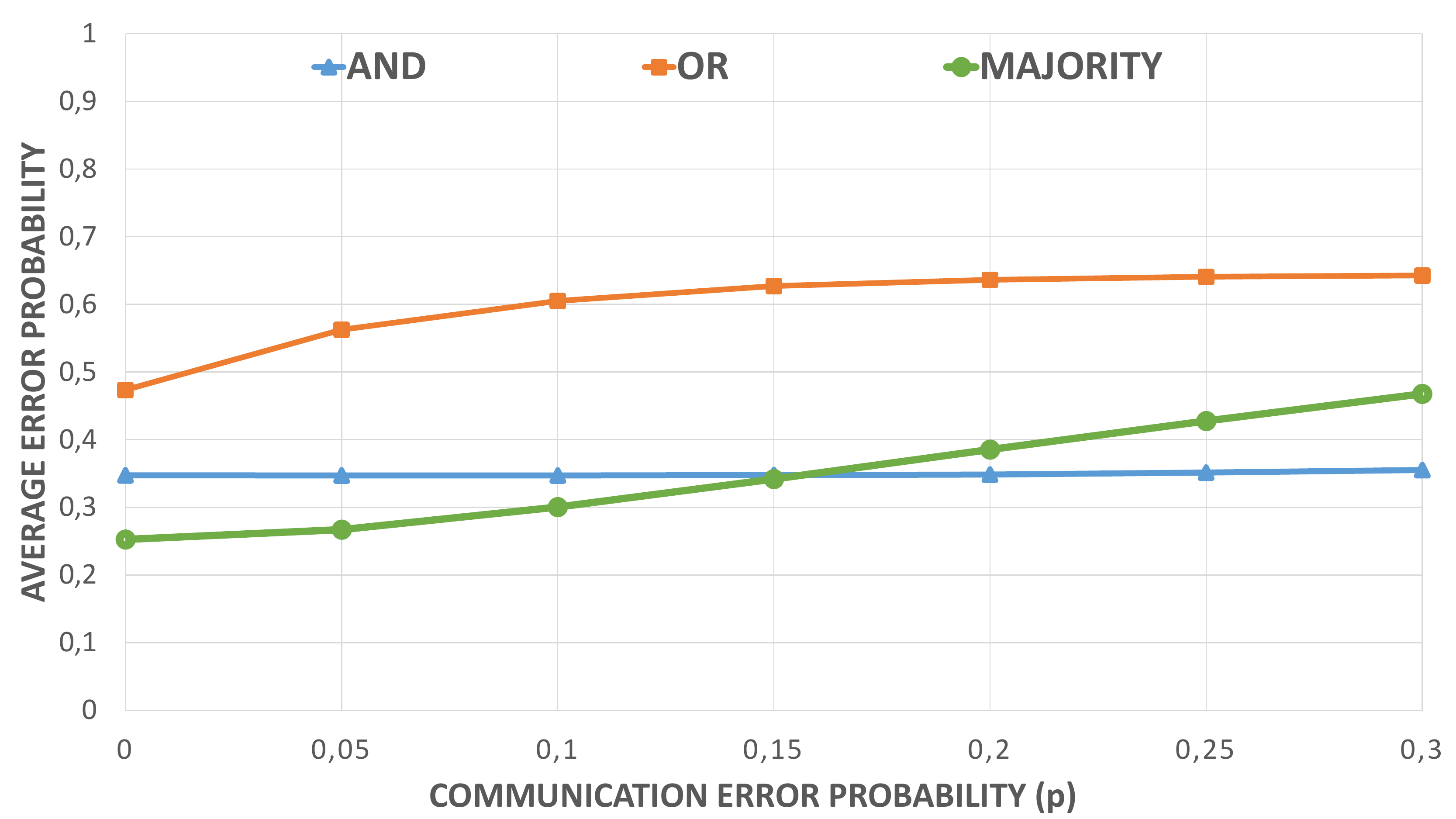}
	\caption{Average Error Probability $P_\textup{er}$ as a function of the communication error probability $p_1=p_2=p$ assuming AND, OR and MAJORITY decision rules for $\gamma = 0.6$ kW and $P_\textup{th}=7.5$ kW. Each point is obtained using Monte Carlo simulation ($10^3$  snapshots).}
	\label{fig:channel-error}
	\vspace{1ex}
\end{figure}

In Fig. \ref{fig:channel-error}, we analyze the effects of the communication error probability on the average error probability assuming that $p_1=p_2=p$, $\gamma = 0.6$ kW and $P_\textup{th}=7.5$ kW.
The first interesting conclusion from the curves is that, even when $p=0$ (error-free), $P_\textup{er}$ assumes a somewhat high value (about $25\%$) even in its best case, which is given by MAJORITY.
When the communication error $p$ increases, $P_\textup{er}$ also increases for OR and MAJORITY while is kept (approximately) constant for AND.
As discussed before, this happens due to the nature of the AND rule, whose performance is determined by the frequency that $s[n]=0$ occurs and the susceptibility of MAJORITY to more frequent communication errors.

All in all, these results indicate that it is possible to get a reasonable average error probability if the event $\sum P_{i,j} > P_\textup{th}$ occurs with low frequency and, therefore, $s[n]=1$ is rare.
In this scenario, AND gate can achieve a low $P_\textup{er}$ since if favors the state $\theta[n]=0$; the drawback is that by choosing such a rule, the decision is weakly related to the system state.
In other words, using AND leads to a quasi-constant guess of $\theta[n]=0$ (regardless of the error events and the actual individual state) so, as the actual system state is $s[n]=0$ anyway, the average error probability tends to be low.
MAJORITY rule in turn better captures the system dynamics, but at the same time is much more vulnerable to communication errors.

\section{Discussions and final remarks}
\label{sec:final} 

In this paper, we analyzed different ways that a wireless sensor network can be implemented using different logical decision rules AND, OR and MAJORITY for a non-critical smart grid application.
Our goal was to show whether it is possible to build a low cost  communication network using only a $1$-bit data signaling. 
We show it is actually possible to attain a low error probability using the proposed scheme if the design parameters are properly set: AND decision rule when the event under consideration is rare.
We also pointed out the weakness of this scheme, which favors the more frequent state and it can be seen as ``always guess that the system is in the more frequent state'' decision rule.
The MAJORITY rule, on the other hand, better captures the system dynamics while it has the drawback of being more susceptible to error events in the communication.

All in all, although the results presented here have some limitations, it clearly opens up new research possibilities.
For example, using different ways of signaling considering more realistic modulations (e.g. Quadrature Amplitude Modulation) and channel models is a good way to have a more robust communication system that is simple and easy to implement.
Other possibility is to use different decision rules like $K$-OUT-OF-$N$, which is a more flexible version of MAJORITY.

Another promising way to develop the proposed framework is to statistically study the average power demand signal.
Our idea is to build a signal processing technique that makes use of the signal statistics, which has been shown it is not Gaussian but rather Weibull or Log-Normal \cite{munkhammar2014characterizing,Sajjad2015}.
%

\bibliographystyle{IEEEtran}


\end{document}